\input{aipcheck}

\pdfoutput=1 

\documentclass[
    ,final            
  ]
  {aipproc}

\usepackage{amssymb}

\layoutstyle{8x11double}

\def\halpha{{H$\alpha$}}
\def\gray{{$\gamma$-ray}}
\def\kms{{km\,s$^{-1}$}}
\def\msun{{M$_\odot$}}
\def\net{{$n_{\rm e} t$}}

\def\netcgs{{cm$^{-3}$s}}
\def\arcmin{{'}}

\def\bar{\overline}

\begin{document}

\title{Multiwavelength Signatures of Cosmic Ray Acceleration by
Young Supernova Remnants}



\author{Jacco Vink}{
  address={Astronomical Institute Utrecht, Utrecht University, P.O. Box 80000, 
 3508TA Utrecht, Netherlands}
}

\begin{abstract}
An overview is given of multiwavelength observations of young supernova 
remnants, with a focus on the observational signatures
of efficient cosmic ray acceleration.
Some of the effects that may be attributed to efficient cosmic
ray acceleration are the radial magnetic fields in young supernova remnants,
magnetic field amplification as determined with X-ray imaging spectroscopy,
evidence for large post-shock compression factors, and low 
plasma temperatures, as measured with high resolution optical/UV/X-ray
spectroscopy. 
Special emphasis is given to spectroscopy of post-shock plasma's, which offers
an opportunity to directly measure the post-shock temperature.
In the presence of efficient cosmic ray acceleration the post-shock
temperatures are expected to be lower than according to standard equations
for a strong shock. For a number of supernova remnants this seems
indeed to be the case.
\end{abstract}

\maketitle


\section{Introduction}

Young supernova remnants are interesting from many different points of view:
they are important for studying the freshly synthesized supernova material,
thereby giving clues about the supernova explosion itself 
\citep[][for some recent examples]{renaud06,badenes08b}, 
and they are of interest for many physical processes associated with
low density plasmas,
for example collisionless shock physics and non-equilibrium ionization.
However, in keeping with the theme of this symposium, I will focus
here on those multiwavelength aspects that reveal something about the
cosmic ray acceleration properties of young supernova remnants (SNRs).

For a long time SNRs have been considered the prime candidates for
providing the bulk of the cosmic rays observed on earth in the energy range
up to $E \sim 3\times 10^{15}$~eV (the ``knee'' in the cosmic ray spectrum), 
but possibly even up to $10^{18}$~eV (the ``ankle''). 
The supernova energetics and rates 
are sufficient to explain the energy density of Galactic
cosmic rays, but the problem was that the conditions in SNRs
seemed not right for accelerating particle up to, or beyond, energies of
$10^{15}$~eV.
In particular, the magnetic field strengths and turbulence
needed for fast acceleration,
were considered unlikely to exist in SNRs \citep{lagage83}. 

Since the 1950-ies it is known
that at least some particles get accelerated up to relativistic energies,
since SNRs turned out to be sources of synchrotron radio emission, 
which indicated the presence of relativistic
electrons \citep[][for an early discussion]{shklovsky68}. However,
the electrons producing radio synchrotron emission have energies in the
GeV range, far short of the $10^{15}$eV needed to explain the cosmic ray
spectrum.

From the 1950-ies to the mid 1990-ies not much observational progress
was made concerning cosmic ray acceleration by SNRs.
Disappointingly,
\gray\ experiments like COS-B, or EGRET revealed little evidence for
the presence of high energy cosmic rays inside SNRs, although EGRET did
find a possible connection between \gray\ sources and old SNRs near
molecular clouds \citep{esposito96}.

Over the last decade the situation has changed considerably. 
X-ray observations have shown that a) electrons are accelerated to 
$> 10$~TeV energies
in some SNRs \citep{koyama95}, 
b) magnetic fields in SNRs are much larger ($\gtrsim 100$~$\mu$G) than the
compressed average field in the interstellar medium ($\sim 5$~$\mu$G)
\citep[e.g.][]{vink03a,berezhko03c,bamba05,ballet05}.

The presence of $>10$~TeV electrons reveals that SNR shocks are
capable of acceleration particles to very high energies, although it
does not yet provide direct evidence that ions, which constitute
the bulk of the cosmic rays observed on earth, are accelerated 
beyond $10^{15}$~eV.

However, the relatively high magnetic field at least indicates the conditions
for accelerating particles up to $10^{15}$~eV are present. Moreover,
the high magnetic fields probably arise through plasma wave generation by
cosmic rays themselves \citep[e.g.][]{bell04}. 
High magnetic fields in SNRs,
therefore, provide indirect evidence for efficient
cosmic ray acceleration.

Direct evidence that, indeed, ions are accelerated to high energies 
by young SNRs comes from TeV \gray\ observations. 
Several SNRs have now
been detected by Cherenkov \gray\ telescopes 
\citep{aharonian01,aharonian04,aharonian05,albert07}, but the interpretation
of the results are hotly debated \citep{butt08}: 
is the \gray\ emission caused by
inverse Compton scattering of $> 10$ TeV electrons, or is it due to 
the decay of neutral pions created by energetic ion-ion collisions?

Although this is an ongoing debate, we will likely make major
advances in our understanding in the next decade.
One of the reasons is the rapid developments in TeV astronomy,
but multiwavelength data are necessary to interpret the TeV data
(e.g. to estimate densities, magnetic fields), and as I will show 
multiwavelength observation will help to reveal
whether ions are efficiently accelerated, 
where and when they will are accelerated,
and how much of the shock energy will be taken up by cosmic rays.

\section{Supernova remnant shocks}
\subsection{Shock evolution}

A typical supernova explosion blows material into the 
interstellar/circumstellar medium (CSM/ISM)
with velocities up to 10,000-20,000 \kms,
and a total kinetic energy of $\sim 10^{51}$~erg. This ejected material
is initial hot from the supernova explosion itself, 
but due to the high expansion rate,
the ejecta cool very fast, even allowing dust to condense (as detected in
the SNR Cassiopeia A \citep{lagage96,rho08}). 
The outer ejecta drive a shock wave
through the CSM/ISM, heating the plasma to temperatures exceeding $10^7$~K.

The hot plasma resulting from the shock is at the outside bound by unshocked
CSM/ISM, but at the inside by fast moving, cool ejecta. The high pressure
in the shocked medium starts driving a shock wave back into the ejecta, thereby
heating the ejecta. This shock is called the 
{\em reverse shock}. Note that the direction of the reverse shock in the frame
of the observer depends on the evolutionary phase of the supernova remnant:
initially the reverse shock is also moving outward at 
high speed, and the faster moving ejecta ``bump'' into the reverse shock. 
At a later stage, occurring when the swept up CSM/ISM mass exceeds several
times the ejecta mass, the reverse shock moves in an opposite
direction to the forward shock, i.e. it moves toward the center. 
Once it reaches the center, all ejecta have been shock heated and the core
of the SNR is hot.
Such a SNR can no longer considered to be a {\em young} SNR, although
some SNRs in this stage of the evolution still show evidence 
for highly metal enriched ejecta
\citep[e.g. the LMC remnant Dem L71][]{vanderheyden03,hughes03}.

\begin{figure}
  \includegraphics[angle=-90,width=\columnwidth]{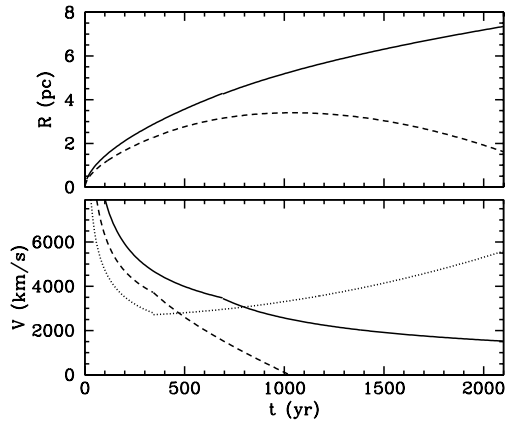}
  \caption{Shock radii and shock velocities as a function of time
according to the Truelove-McKee $n=7, s=0$\ model \citep{truelove99}.
The solid line represents the forward shock, whereas the dashed line represents
the reverse shock. In the bottom panel the reverse shock velocity in
the frame of the ejecta is shown as a dotted line.
The dimensionless model was adjusted to fit the properties of Kepler's 
SNR \citep{vink08b}.
\label{fig_truelove}
}
\end{figure}

There are several analytic models for the shock structure and evolution 
of SNRs \citep{chevalier82,truelove99,laming03}. Fig.~\ref{fig_truelove}
shows the evolution of a SNR in a uniform ISM, as may be appropriate for
Type Ia SNRs, which are thought not to have a strong stellar wind prior to
explosion. The figure illustrates the behavior of the reverse shock,
whose radius peaks, in this particular case, at around 1000 yr.

The asymptotic behavior of the shock radius is that of the Sedov-Taylor 
self-similar evolution:i.e. :
\begin{equation}
R = \Bigl(2.03 \frac{E t^2}{\rho_0} \Bigr)^{1/5},\,  
V_s = \frac{dR}{dt} = \frac{2}{5} \frac{R}{t},
\end{equation}
with $E$\ the explosion energy, and $\rho_0$ the ISM density.

The medium surrounding a SNR can be rather complex. In particular,
around core collapse supernovae one expects several regions shaped by
different evolutionary stages of the progenitor. On the main sequence a 
massive star blows a fast ($v_w \sim 1000$ \kms), 
tenuous wind, but in the red supergiant phase a massive star loses mass
with a rate of $\sim 10^{-4}-10^{-5}$~\msun, in a slow ($v_w \sim 10$ \kms)
dense wind. Finally, the most massive stars may become Wolf-Rayet stars, 
which again have a very fast, tenuous wind. As a result the CSM may consist
of several layers separated by shells caused by the interactions of winds
from different phases in the star's life.

If one considers an SNR evolving in a red supergiant (RSG) phase, which may
be quite common, one expects the following density profile \citep{chevalier82}:
\begin{equation}
\rho_{CSM} = \frac{\dot{M}}{4\pi r^2 v_w}. \label{eq_wind}
\end{equation}
The young SNR Cas A may be currently in this phase 
\citep{chevalier03,schure08}. The shock velocity in such a wind structure
quickly follows the relation $V_s = \frac{2}{3} \frac{R}{t}$, 
which is indeed
close to what is found for Cas A \citep{vink98a,delaney03}.

\subsection{Shock heating and temperature equilibration}

\begin{figure}
\includegraphics[angle=-90,width=\columnwidth]{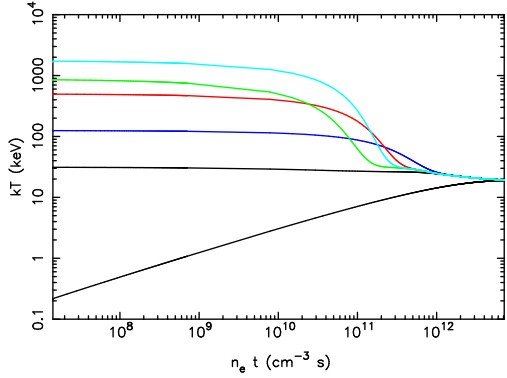}
\caption{
Theoretical model of the collisional equilibration process for a plane
parallel shock model. It is assumed that the shock velocity is 
$V_s = 4000$~\kms. The equilibration scales with the product of density and
time, parameterized here as $n_{\rm e} t$. The different lines indicate the
temperature of different particle species. From bottom to top (on the left):
electrons, protons, helium, oxygen, silicon, and iron.
\label{fig_equil}
}
\end{figure}

The equations of mass-, momentum- and energy-flux conservation
gives the following relation between shock velocity, $V_s$,
and temperature for a high Mach number shock
\citep[e.g.][]{zeldovich66}:
\begin{equation}
k\bar{T} = \frac{3}{16} \mu m_{\rm p} V_s^2.\label{eq_hugoniot1}
\end{equation}
$\bar{T}$ refers here to the average temperature, i.e.
the average kinetic energy per particle. 
However, the microphysics of shock heating is not well known.
SNR shocks are collisionless, which means that particle-particle collisions
(Coulomb interactions) are rare, 
and insufficient to heat the plasma within a time comparable to
the age of the SNR. Instead, 
the heating probably occurs through plasma waves. In such
a case it is not quite known whether different particle species are heated to
the same temperature. Alternatively, one may expect that all particle
species have the same velocity distribution. This could arise if plasma
waves in the shock scatter the incoming particles in different directions,
with only small changes to 
the absolute velocities of the particles, which is $V_s$ for
all particles. In such a case the temperature for each particle
species is different and scales with the mass, $m_i$, of the particle:
\begin{equation}
kT_i = \frac{3}{16} m_i V_s^2. \label{eq_hugoniot2}
\end{equation}
In the post-shock region, full equilibration will take place on a time scale
of $\log (n_{\rm e t}) \approx 12.5$ \citep{itoh77}, 
corresponding to $\sim 30,000$~yr of a typical
density of $n_{\rm e} \approx 1$~cm$^{-3}$ (Fig.~\ref{fig_equil}).
Note that equilibration by particle-particle interactions depends on the charge
of the particles and the mass ratio. For that reason electron-proton 
equilibration takes a long time, whereas proton-iron equilibration proceeds
on a time scale of $\log (n_{\rm e t}) \approx 11.5$. 
The parameter \net\ 
is also important for the ionization process, and can be
directly measured using X-ray line ratios.\footnote{I will skip the details
here, but for those interested, more information and background
can be found in \citep{itoh77,gronenschild82,hughes85,borkowski01b,vink05b}.
} 

Electron temperature can be determined using X-ray spectroscopy,
even employing CCD detectors. 
Ion temperatures can only be determined by measuring the thermal Doppler
broadening of spectral lines. In order to eliminate Doppler
broadening caused by line of sight motions due the SNR expansion, 
spectra should be obtained of the rim of a SNR, 
where only motions in the plane of the
sky are to be expected. 
Most measurements of ion temperatures concern the hydrogen lines.
Since hydrogen is quickly ionized, the hydrogen line emission
arises from very close to the shock front. The presence of hydrogen lines
is only possible, if some neutral hydrogen is present in the ISM/CSM.
The direct excitation in the post-shock gas gives rise to a narrow line, as
the neutral hydrogen has not yet interacted with the shock heated plasma,
and still has the temperature of the CSM. 
Charge exchange
between an incoming neutral hydrogen atom and a proton in the heated gas,
gives rise to a broadened line. The width of this broad component is a direct
measurement of the proton temperature behind the shock
\citep{chevalier80,heng07}. 
In the case of equilibration of
temperatures, the proton temperature is equal to the mean plasma temperature.

X-ray spectroscopy, in general, prodives a direct measure of the electron 
temperature. The electron temperature, together with \net, determines
the relative line ratios of lines of a given ion or atom. 
Also the characteristic cut-off
in the bremsstrahlung continuum is directly related to the electron 
temperature. In young SNRs, however, synchrotron radiation is another
source of continuum radiation, and the two mechanism are not always easy
to tell apart. 

In principle, in X-rays one can also measure thermal Doppler
broadening, but in practice the CCD detectors on board current X-ray 
observatories like Chandra and XMM-Newton
pair great imaging capabilities with too poor spectral resolution to measure
thermal Doppler widths. The  high resolution grating spectrometers on
board Chandra and XMM-Newton have a good spectral 
resolution, but they are slitless and the spectral quality is degraded
for extended sources.
Nevertheless, X-ray Doppler broadening has been
measured for SN1006 using XMM-Newton's Reflective Grating Spectrometer 
\citep{vink03b}.
This remnant has one of the lowest values of \net:
\net$\approx 2\times 10^{9}$~\netcgs.
Optical, UV and X-ray spectroscopy of the northwestern part of this
remnant reveals that ions, protons and electrons are out of temperature 
equilibrium \citep{ghavamian02,raymond95,vink03b}, with the electrons
being much cooler than the oxygen ions 
($kT_{\rm e}\approx 1.5$~keV, $kT_{\rm O VII} \approx 500$~keV 
\citep{vink03b}).

Apart from SN1006, also in several other SNRs have ion temperatures
been measured, but only using optical and UV lines 
\citep{smith94,ghavamian01,ghavamian07,ghavamian07b}.
Combining these measurements with electron temperatures obtained from
X-ray spectroscopy shows that for the fastest shocks ($V_s > 1000$~\kms)
the electron temperature
can be as low as 10\% of the proton temperature. For the slow shocks, like
in the Cygnus Loop, electrons and ions seem to be in full equilibrium, with 
a turnover from equilibrium to non-equilibrium occuring around
$600$~\kms \citep{ghavamian07}. Toward the end of this paper I will return
to the issue of ion temperatures in the context of cosmic ray acceleration.

\begin{figure*}
\includegraphics[viewport=18 -20 575 487,clip=true,width=0.5\textwidth]{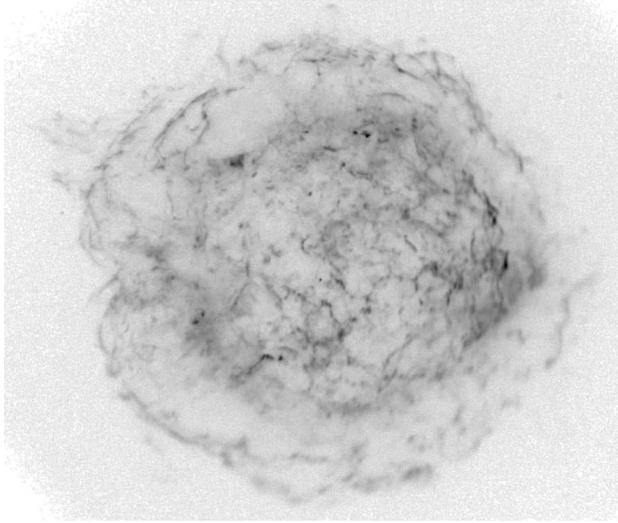}
\includegraphics[viewport=11 142 591 692,clip=true,width=0.5\textwidth]{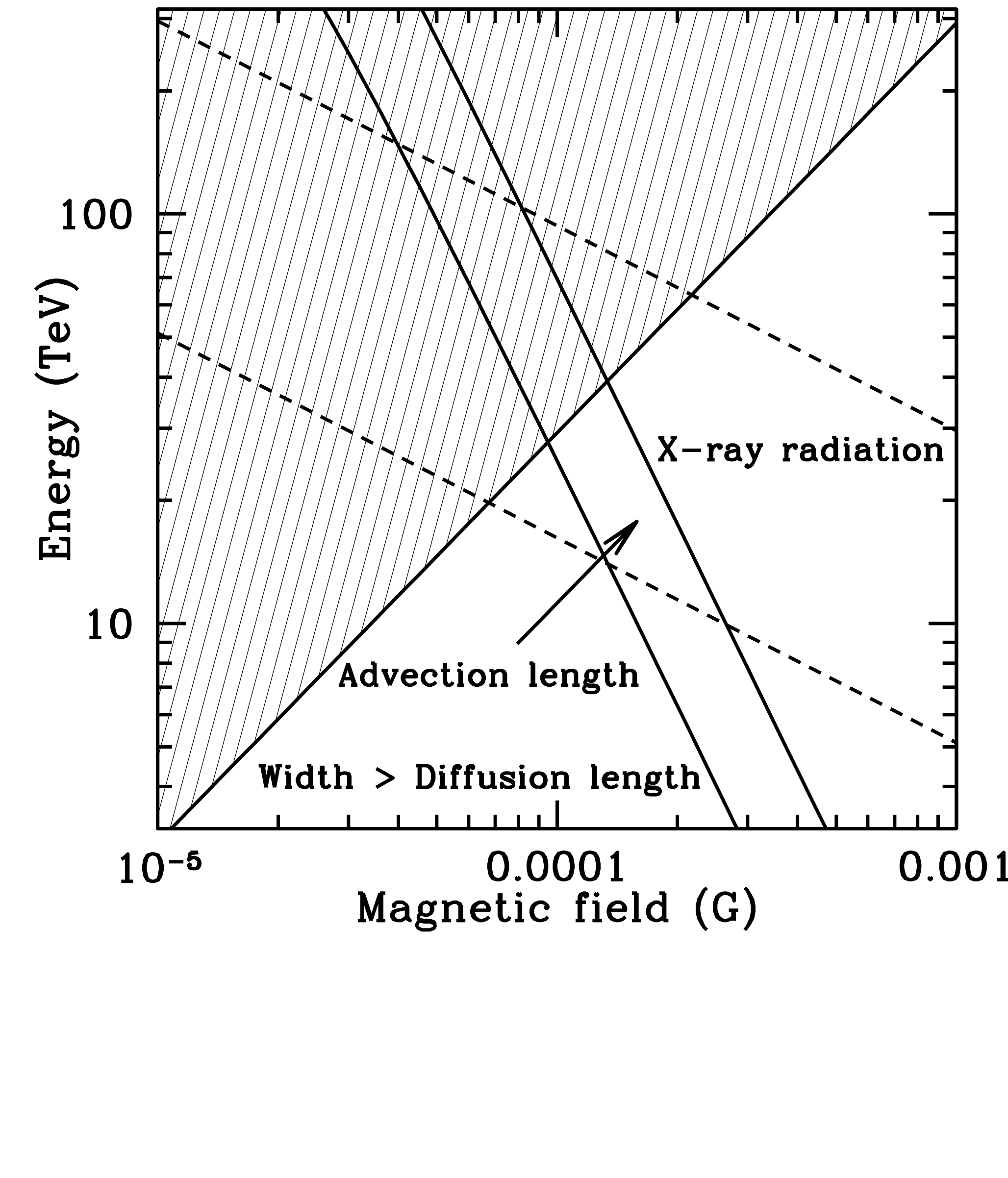}
\caption{
A deep Chandra image of Cas A \citep{hwang04}
in the 4-6 keV continuum band (left).
Note the thin filaments, marking the border of the remnant
The remnant has a radius of about 2.5\arcmin.
Right: The maximum cosmic ray electron
energy versus magnetic field strength
for the region just downstream of Cas A's shock front, as determined
from the thickness of the filaments. The shaded area
is excluded, because the filament width cannot be smaller than the minimum
possible diffusion length \citep[c.f.][]{vink03a}.
(These figures were published before in \citep{vink06b}.)
\label{fig_casa}
}
\end{figure*}

\section{Multiwavelength imprints of efficient cosmic ray acceleration}

The coming of age of TeV and X-ray astronomy over the last 15 years,
has greatly increased our knowledge of cosmic ray acceleration by
SNRs. Here I will explain some of the observations that form the
basis of our current understanding. Not included here is TeV astronomy,
which is amply addressed in other parts of these proceedings.

\subsection{Measuring magnetic fields using X-ray synchrotron rims}
The discovery by the ASCA satellite 
that the X-ray continuum emission from SN1006 is dominated
by synchrotron emission \citep{koyama95},\footnote{This was
anticipated by Reynolds \& Chevalier \citep{reynolds81}.} 
has been the start of a number of
discoveries in X-rays, 
which all relate to the cosmic ray acceleration properties of SNRs.

One of these discoveries is that the X-ray synchrotron emission from SNRs
is confined to a region very close to the shock front.
In the SNRs Cas A, Kepler (SN1604) and Tycho (SN1572), the synchrotron
emission comes from a region only a few arcseconds near the shock front
\citep{gotthelf01a,hwang02}. In fact, the identification of the X-ray
synchrotron emission required the superior angular resolution of Chandra.
In some other remnants, SN1006, RCW 86, RX J1713.7-3946, and RX J0852.0-4622,
the X-ray synchrotron emission comes from a larger region, although some
fine scale structure is present.

The size of the X-ray emitting region is now generally believed to
be determined by the magnetic field strength. The lifetime of
a relativistic electron in a magnetic field is in cgs units:
\begin{equation}
\tau_{loss} = \frac{637}{B_\perp^2E} \, {\rm s}.
\end{equation}
The typical photon energy $E_{ph}$ corresponding to an electron energy $E$ is:
\begin{equation}
E_{ph} = 7.4 E^2B_\perp \, {\rm keV}.
\end{equation}

The electrons are accelerated at the shock front by diffusive shock 
acceleration. At small scales their trajectories are determined by scattering
on plasma waves (diffusion), but on large scales their average motion follows
the plasma. If the shock compression ratio is 
$\chi \equiv \rho_0/\rho_1$, then the plasma velocity with respect
to the shock is $V_1 = V_s/\chi$. As the electrons are moving slowly
away from the shock front, they lose energy, until they are no
longer emitting X-ray radiation. Thus, the loss time corresponds to the width
of the X-ray emitting region through:
\begin{equation}
l_{loss} =  \tau_{loss} \frac{1}{\chi}V_s.
\end{equation}
By combining the measured width and the measured/inferred shock velocity
with the observed photon energy one arrives at an estimate of the post-shock
magnetic field strength \citep{vink03a}, $B$, as illustrated
in Fig.~\ref{fig_casa}.

Several groups \citep{bamba03b,berezhko03c,berezhko03a} 
do not use the loss length scale, $l_{loss}$, but instead
assume that the length scale seen corresponds to the
diffusion length scale, $l_{diff}$, 
i.e. the length scale at which diffusive motions are
more important than the bulk plasma motion. However, in practice this
method should give similar results, as long the synchrotron emission
comes from electrons near the maximum acceleration energy, where diffusive
acceleration is balanced by radiative losses, since $l_{diff} \leq l_{loss}$
for efficient acceleration
\citep{vink05b,parizot06,vink06d}.\footnote{
Interestingly, the two method have different assumptions: to estimate
$l_{diff}$ one has to assume Bohm-diffusion, whereas for
estimating $l_{loss}$ one has to assume a compression ratio. The fact that
both methods give similar magnetic field values is an indication
that both assumptions must be approximately valid \citep{vink05b}.
}

Magnetic field measurements of several SNRs have now been published, 
and indicate post-shock magnetic fields of $20-600$~$\mu$G 
\citep{vink03a,berezhko03c,berezhko04a,voelk05,bamba05,ballet05,warren05,vink06b}.
\begin{figure}
\includegraphics[angle=-90,width=\columnwidth]{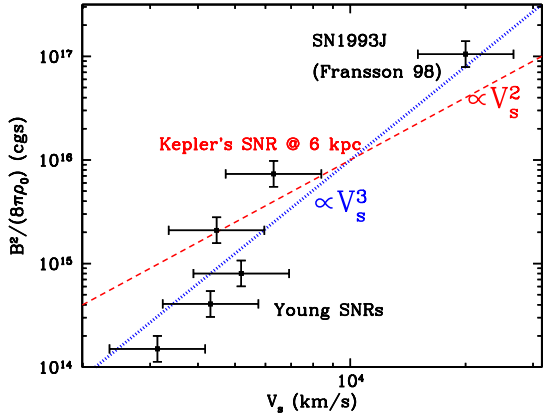}
\caption{
Dependence of the 
post-shock magnetic field energy density $B^2/(8\pi \rho_0)$
on the shock velocity $V_s$. The dashed line shows a $V_s^2$ dependency
and the dotted line a $V_s^3$ dependency.
The input values can be found in \citep{vink06b}, 
except for SN1993J \citep{fransson98} and Kepler's SNR, 
for which the shock velocity in the southwest
was taken from \citep{vink08b} and the density from \citep{cassam04}.
\label{fig_bfield}
}
\end{figure}
This is higher than expected based on the compression of the average magnetic
field in the ISM, $B\sim 5$~$\mu$G, and, therefore, suggests that some
form of magnetic field amplification mechanism is operating.
The most likely mechanism is non-linear growth of plasma waves
induced by streaming of cosmic rays 
\citep{bell01,bell04,zirakashvili08b}, but some alternative, perhaps
complementary, mechanisms have been proposed 
\citep{bykov05,giacalone07}.

Recently, high magnetic fields for SNRs were also reported based
on small X-ray synchrotron brightness fluctuations in Cas A
and RX J1713.7-3946 \citep{patnaude07,uchiyama08,uchiyama07}.
These findings support the presence of relatively
high magnetic fields in young SNRs.

It is not yet clear what the relation is between the
ram pressure at the shock front, $\rho_0 V_s^2$, and the post-shock
magnetic field pressure. Based on observations it has been argued
that $B^2 \propto \rho_0 V_s^2$ \citep{voelk05}, whereas in \citep{bell04}
it has been argued that  $B^2 \propto \rho_0 V_s^3$.
In the latter case, a proportionally larger fraction of the incoming
kinetic energy is transferred to cosmic rays and magnetic fields for
during the earliest life of an SNR, since then
the shock velocity is highest.

The problem with the current measurements of SNRs is that the dynamic
range in density is quite high, but the dynamic range in shock velocity is
small, since the magnetic fields can only be measured for X-ray synchrotron
emitting remnants, and only SNRs with $V_s \gtrsim 2000$~\kms 
\citep{aharonian99}  are expected to emit X-ray synchrotron radiation,
whereas most Galactic remnants known have $V_s < 5500$~\kms.
However, for supernova SN1993J, 
which has a shock velocity $\sim 20,000$~\kms, a high magnetic field of
64~G has been reported \citep{fransson98}. If one assumes that for
this object the same mechanism is at work, one may try to distinguish
between the two magnetic field strength scalings. As Fig.~\ref{fig_bfield}
shows, including SN1993J favors a magnetic field strength scaling
as $B^2 \propto \rho_0 V_s^3$, but one should regard this with some
caution, since the magnetic field in 
SN1993J may have different origin than the magnetic fields in young SNRs.

\begin{figure}
\includegraphics[width=\columnwidth]{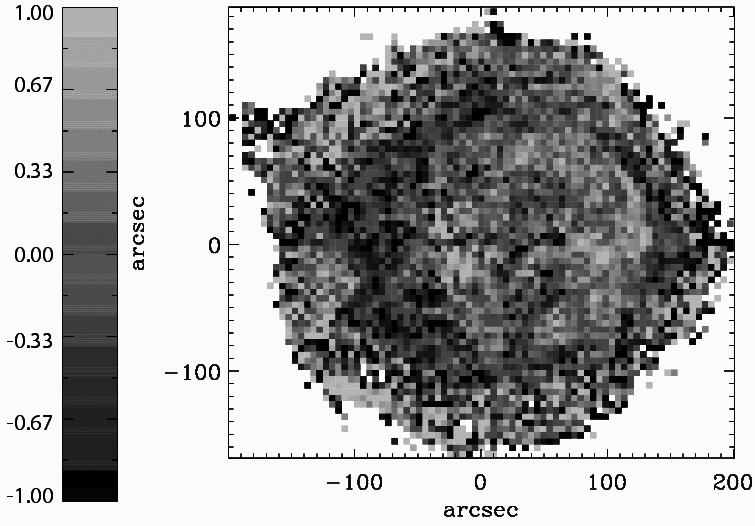}
\caption{
X-ray continuum spectral index map of Cas A based on the 1~Ms exposure
by Chandra \citep{helder08}. The gray scale  coding indicates spectral number
index with respect to $\Gamma = -3.1$. A hard spectral index (lighter pixels) 
is an indication for X-ray synchrotron emission. 
One sees here that hard spectral indices at the rims and in the central region,
somewhat shifted to the west of the SNR.
(Figure made by Eveline Helder, a color version can be found in
\citep{helder08}.)\label{fig_helder}
}
\end{figure}
\subsection{Magnetic field amplification at the reverse shock}

The high magnetic field of SN1993J raises the question how universal
magnetic field amplification is. Does it require certain preconditions,
like a medium strength magnetic field, to start with?

Interestingly, it was recently reported that for the young SNR Cas A
a large part of the X-ray synchrotron emission is coming from the
reverse shock, in particular in the western region of Cas A 
\citep{helder08,uchiyama08} (Fig.~\ref{fig_helder}).

In the past, reverse shocks as sites of particle acceleration were often
ignored (but see \citep{rho02,ellison05}), since it was assumed that
the magnetic field in the ejecta was low, due to their large scale expansion.
However, particle acceleration at the reverse shock, sufficiently fast
to give rise to X-ray synchrotron emission, may be important for two reasons:
1) it shows that magnetic field amplification is rather universal, and that it
does not require relatively large initial magnetic fields 
2) some cosmic rays may be accelerated from metal rich material, which could
lead to signatures in the cosmic ray composition. Indeed there is some evidence
that 20\% of the Galactic cosmic rays come from massive stars
\citep{wiedenbeck81,binns08}.

As remarked by Luke Drury at the symposium, one should not
overestimate the significance of the reverse shock for their contribution
to the observed cosmic ray spectrum:
cosmic ray acceleration is efficient when the physical
shock velocity is high. For the reverse shock this occurs
when the shock starts to
move back toward the center of the SNR, until it reaches the center. This
is a relatively short period in the life of a SNR. Moreover, during this
time the area spanned by the reverse shock is  smaller than that
of the forward shock, so the total number of particles
entering the reverse shock is much less than the
number of particles swept up by the forward shock. 
Both these aspects are illustrated in 
Fig.~\ref{fig_truelove}. For Cas A it is believed that the reverse shock
velocity in the frame of the freely expanding ejecta is less than
$2000$~\kms \citep{laming03}, except in the western region, where somehow
the reverse shock is almost at a standstill in our frame, and the ejecta are
shocked with the ejecta
free expansion velocity of $4000-8000$~\kms \citep{helder08}.

\subsection{What is the maximum energy  cosmic rays can be accelerated to?}

Several authors have pointed out that the high magnetic fields that have
been inferred from X-ray synchrotron emitting rims indeed allow protons to
be accelerated to $>10^{15}$~eV, and heavy ions can in principle accelerated
to even higher energies \citep[e.g.][]{vink05b,parizot06}.
It is also clear that the magnetic field energy density scales with:
\begin{equation}
B^2 \propto \rho_0 V^{\alpha}, \label{eq_bfield}
\end{equation}
with $\alpha =2$ or $3$.
As a consequence higher magnetic fields are present early in the life
of a SNR when the shock velocity is higher. This has led to the suggestion
that the highest energy  cosmic rays are accelerated early on,
and, as the magnetic field drops, those particles escape first.
The peak energy is thus a function of time. The observed cosmic ray
spectrum is in such a scenario a superposition of cosmic rays released
over an extended period of time \citep{ptuskin05}

Note that not only does the magnetic field energy depend on $V_s$, but
also on $\rho_0$. 
This means that the highest energies are reached for shocks in dense 
environments. Some supernovae have by their very nature high 
circumstellar densities.
As indicated by Eq.~\ref{eq_wind}, 
the density around a massive, wind blowing star, 
is highest for slow wind speeds, $v_w$, and high mass loss rates.
So it is likely that the highest cosmic ray energies are obtained by SNRs
developing in a RSG wind. An example, as mentioned above, is
Cas A. In fact, an optical spectrum of the light echo of the supernova 
explosion
reveals that it is the remnant of a Type IIb supernova, very similar
to the bright radio supernova SN1993J \citep{krause08}! 
Cas A's progenitor probably had only a very short Wolf-Rayet star
phase, if any at all \citep{chevalier03,young06,schure08}.

An SNR like Cas A reaches the self-similar shock evolution very early
on, in which case $R_s \propto t^{2/3}$ and $V_s \propto t^{-1/3}$\
\citep{chevalier82}. One can use
this similarity evolution to estimate the maximum proton cosmic ray
energy as a function of time: 
First note that $E_{max} \propto t_{acc} B_1 V_s^2$, with
$t_{acc}$ the acceleration time, for which we can take the age of the SNR.
Using Eq.~\ref{eq_wind} and Eq.~\ref{eq_bfield} one finds:
\begin{equation}
B_1 \propto \frac{1}{R_s} t^{-\alpha/6}  \propto t^{-2/3 - \alpha/6}.
\end{equation}
We therefore find for the maximum proton energy in a young SNR evolving
in a dense stellar wind:
\begin{equation}
E_{max} \propto t^{1 - 4/3 -\alpha/6}, \label{eq_emax}
\end{equation}
for $\alpha =3$ this means $E_{max} \propto t^{-5/6}$, and for $\alpha=2$
this is $E_{max} \propto  t^{-2/3}$.

The value for the magnetic field found for Cas A ($B\approx 0.5$~mG) 
is sufficient for $E_{max}$ presently to be $\sim 10^{15}$~eV.
However, Cas A's ability to accelerate cosmic rays was probably even
better in the past: according to Eq.~\ref{eq_emax} and using $\alpha=3$, 
when Cas A was only 30~yr old the maximum proton 
energy was $E_{max} \sim 7\times 10^{15}$~eV and its post-shock
magnetic field strength 7~mG. Note that the evolution of the magnetic
field ensures that the diffusion length scale (assuming Bohm diffusion) 
does not exceed the size
of the remnant, 
since $B$ evolves faster with $t$ than $R_s$. 
The $1/r^2$  density profile makes
that the flux of particles does not depend on the shock radius.

All this implies that probably most of the Galactic cosmic
rays with $E > 10^{15}$~eV were accelerated by Type II/Type IIb
supernova remnants, during the first 100~yr of their lives. Most massive
stars explode while in their RSG phase. Only stars more
massive than 25~\msun\ (a minority), 
probably explode in the Wolf-Rayet stars phase.
Their SNRs are probably less capable of acceleration cosmic rays to
very high energies, unless they explode with larger energies,
as in the case of hypernovae.

\subsection{Measuring compression ratios using X-ray imaging}
Magnetic field amplification is probably a result of cosmic ray streaming.
The presence of relatively large magnetic fields in young SNRs, therefore,
implies by itself efficient cosmic ray acceleration.
I use the word ``efficient''
here in contrast to fast acceleration. Fast acceleration means
that cosmic rays are accelerated to high energies, efficient that
the cosmic ray content of SNRs is energetically important.

Efficient cosmic ray acceleration affects both the evolution of
the SNR itself \citep{decourchelle00}, 
and the cosmic ray spectrum \citep{ellison91,berezhko99,blasi05}.
If cosmic ray acceleration is very efficient the cosmic rays provide
back-pressure to the unshocked plasma, which gives rise to
a concave spectrum, i.e. it is steeper than the test particle spectrum
at low energies and flatter at high energies.
If the internal energy of the post-shock gas is dominated by
{\em relativistic} cosmic rays the compression ratio increases. A standard high
Mach number shock in a monatomic gas ($\gamma=5/3$)
will have a compression ratio of $\chi = (\gamma+1)/(\gamma -1) = 4$,
whereas
for a gas dominated by relativistic particles ($\gamma=4/3$) this is
$\chi = 7$.
Compression ratios in excess of 7 are possible, if energy losses are taken into
account. 

In old SNRs radiative energy losses lead to higher compression ratio,
since shock velocities of $V_s \lesssim 200$~\kms,
produce plasma with a temperature of $<10^6$~keV, near the peak of the cooling
curve. This gives rise to the formation of the filamentary structures radiating
in forbidden lines, which make for beautiful Hubble Space Telescope images,
of SNRs like Vela, the Cygnus Loop or, N49.

For young SNRs, which efficiently accelerate cosmic rays, energy
may leak out of the plasma due to the escape of cosmic rays. The highest energy
cosmic rays are the most likely ones to escape. Escaping cosmic rays
are only energetically important for flat, or concave spectra.
In that
case the highest energy cosmic rays may contain a significant fraction of
the total internal plasma energy.
Although concave spectra are expected in most efficient 
cosmic ray acceleration models,
this is by no means a certainty yet; see the contribution
by V. Zirakashvili in these proceedings.

Higher compression ratios due to cosmic ray escape will give rise to
different ratios of the forward and reverse shock radii
or of the forward shock and contact discontinuity radii 
\citep{decourchelle00}, 
the contact discontinuity being the boundary between shocked supernova ejecta
and shock heated CSM.

Indeed, for both Tycho's SNR \citep{warren05} and SN1006 \citep{cassam08}
it has been reported that the contact discontinuity lies very close
to the forward shock. SN1006, a newly detected H.E.S.S. source, 
is an interesting SNR in this respect, since the X-ray synchrotron
emission, and also most of the radio emission, is confined to
the northeastern and southwestern regions. If we take the shocks in 
these regions sites of efficient cosmic ray acceleration, one may expect
that in those regions the contact discontinuity and forward shock are closer to
each other than in the rest of the remnant, due to
a higher compression ratio. This is indeed what is found,
but surprisingly also in the other regions of SN1006 the contact discontinuity
is closer to the forward shock than expected \citep{cassam08}.

This suggests that apart from a higher compression ratio also some
other mechanisms (e.g. ejecta clumping, hydrodynamical instabilities) 
are at play in bringing the ejecta close to the shock front.
So SNR morphology hints at high compression ratios, signifying efficient
cosmic ray acceleration, but a quantitative results cannot yet be obtained.

\subsection{Radio polarimetry and magnetic field amplification}

Fast cosmic ray acceleration requires a small diffusion constant,
for which, apart from a relatively high magnetic field, also a turbulent
magnetic field is required, i.e. $\delta B/B \sim 1$. Magnetic field
amplification by growth of plasma waves due to streaming instabilities
\citep{bell04}, naturally provides
both the magnetic field strength and its turbulence. Observationally,
turbulent magnetic fields are implied by the fact that near the maximum
electron energies the diffusion length scale and advection (loss time)
length scales are similar (see \citep{vink05b} and above).

\begin{figure*}
\includegraphics[angle=-90,width=0.5\textwidth]{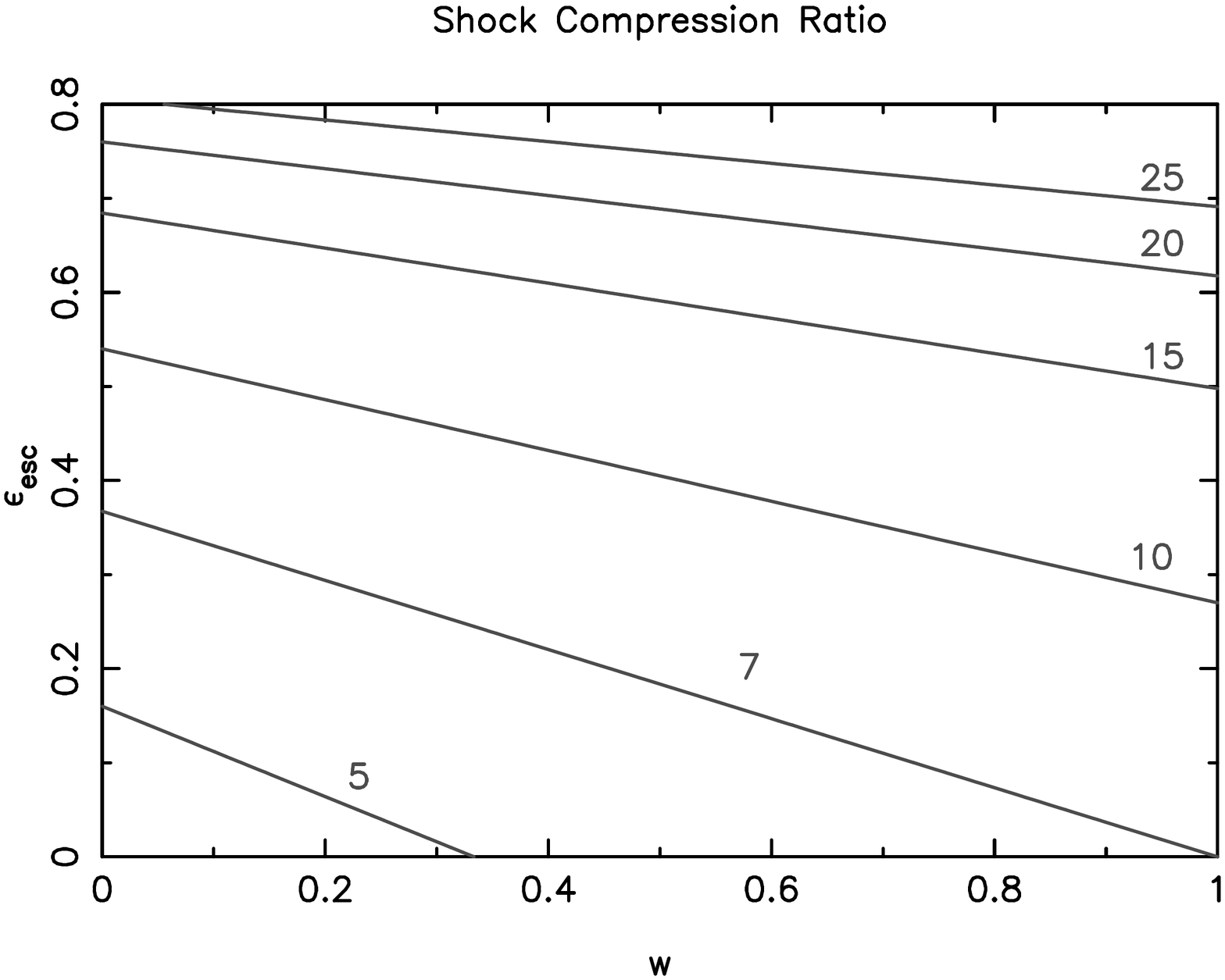}
\includegraphics[angle=-90,width=0.5\textwidth]{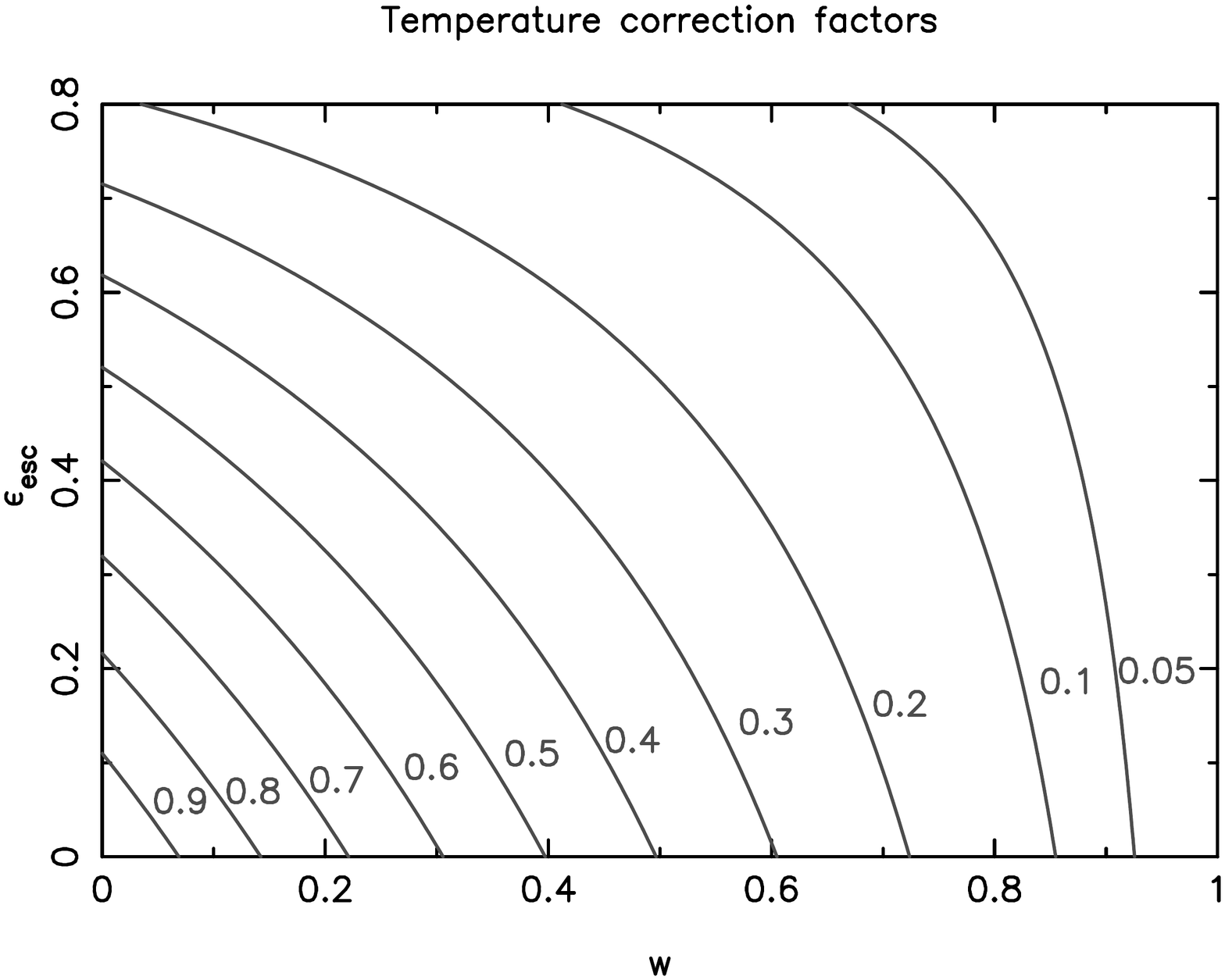}
\caption{
Graphical representation of 
the effects of cosmic ray escape 
and relative cosmic ray pressure ($w$) on the compression ratio (left)
and the post-shock temperature (right). In the right hand figure, the
values are given as correction factors with respect to the standard
strong shock expressions (Eq.~\ref{eq_hugoniot1},\ref{eq_hugoniot2}).
\label{fig_hugoniot}}
\end{figure*}

Other evidence that the magnetic fields are turbulent comes from
the long known observations of radio polarization of SNRs. These
indicate that old SNRs have preferentially tangential magnetic fields,
whereas young SNRs have radial magnetic fields \citep{dickel76}.
Some recent results regarding magnetic field structure can be
found for Cas A \citep{gotthelf01a},
 Kepler's SNR \citep{delaney02}, RCW 86 \citep{dickel01} 
and the old SNR PKS 1209-51  \citep{milne94}.

The tangential magnetic fields in old SNRs are caused by the compression
of the ISM magnetic fields, which have some large scale coherence 
(see the contribution by R. Beck in this volume). Shocks only
compresses the tangential component of the field. In those regions
where the uncompressed magnetic fields are perpendicular to the shock normal,
the post-shock magnetic field will be enhanced, increasing the radio brigthness
of these regions.

The radial magnetic fields in young SNRs are not that easily explained
by MHD simulations \citep{jun96,schure08b}. Rayleigh-Taylor instabilities
at the contact discontinuity, indeed, produce  radial magnetic fields
\citep{jun96}. However, near the shock front, simulations produce tangential, 
not radial magnetic fields. Efficient cosmic ray acceleration,
may indeed be the missing ingredient. This can happen in two ways.
First of all, efficient cosmic ray acceleration gives rise to
turbulent magnetic fields. Since these have no preferred direction,
strong tangential fields where the shock normal is perpendicular to the
magnetic fields are avoided \citep{schure08b}. 
Secondly, enhanced shock compression will
bring the Rayleigh-Taylor instabilities closer to the shock front 
\citep{blondin01}. However, it is not clear whether the compression is 
sufficiently enhanced to indeed produce radial magnetic fields close 
 to the forward shock. Zirakashvili reported at this symposium that
MHD simulations incorporating magnetic field amplification tends to make
the downstream magnetic field stretch in a radial direction 
\citep{zirakashvili08b}.

It, therefore, seems likely that 
the observed radial magnetic field structure in young SNRs
is either facilitated or a direct by-product of efficient cosmic ray
acceleration and magnetic field amplification.

\begin{figure}
\includegraphics[viewport=57 178 577 597,clip=true,width=\columnwidth]{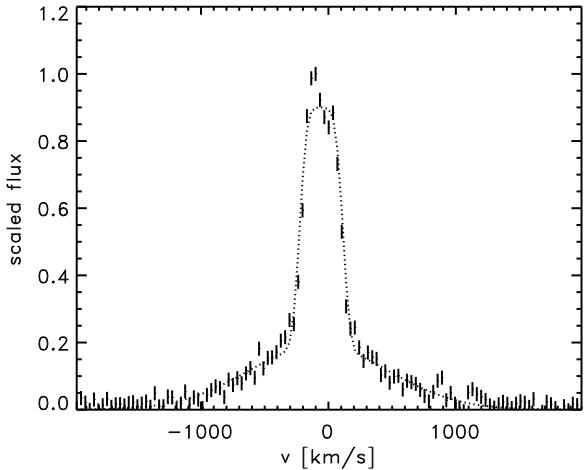}
\caption{
\halpha\ line profile of the northeastern part of RCW 86,
as observed with the VLT (Helder et al. in preparation).
The line consists of a narrow peak, from direct excitation,
and broad wings caused by charge exchange between the incoming neutral
hydrogen atoms and the shock heated protons. The FWHM of the broad wing
is a direct measure of the proton temperature.
The narrow peak is broadend due to the relatively broad line spread function
(LSF) of 330~\kms. 
Preliminary results 
give a FWHM of ($1270\pm 60$)~\kms, after correction for the LSF.
(Figure provided by E. Helder)
\label{fig_rcw86}
}
\end{figure}

\subsection{High resolution spectroscopy and
cosmic ray acceleration efficiency}

In the literature on thermal Doppler broadening of lines from SNRs, one
has concentrated mostly on the equilibration of electrons and ions, 
and the consequences for deriving shock speeds using Eq.~\ref{eq_hugoniot1} or 
 Eq.~\ref{eq_hugoniot2}.
Rarely has the discussion focussed on the influence of cosmic rays
on the measured temperatures.  However, cosmic rays may contain a large
part of the internal energy, and change the adiabatic index
of the plasma. These effects are not taken into account in 
Eqs.~\ref{eq_hugoniot1},  ~\ref{eq_hugoniot2}, which are usually
used to derive shock parameters.

An exception is the case of
the Small Magellanic Cloud remnant 1E 0102.2-7219. As reported in
\citep{hughes00b} the electron temperature of this remnant is too
cold compared to the measured shock velocity, even allowing for
non-equilibration of electron and ion temperatures. A direct measurement
of the ion temperature in this remnant is difficult, as this
oxygen rich remnant does not emit optical hydrogen lines.

How much does efficient cosmic ray acceleration influence the
post-shock plasma temperatures?
The answer depends on two parameters
1) the fraction of the pressure contributed by
non-thermal components, 2) the energy escape fraction associated
with cosmic rays.
For a strong shock one can parameterize these with:
\begin{equation}
w \equiv \frac{P_{NT}}{P_{T} +P_{NT} },
\end{equation}
the relative pressure of the non-thermal components (cosmic
rays and magnetic fields) \citep[c.f.][]{chevalier83,blasi05},
and the shock compression ratios, which for a strong shock is:
\begin{equation}
\chi = \frac{ G + \sqrt{G^2 - (1 -\epsilon_{esc})(2G -1)}}{1 -\epsilon_{esc}},
\label{eq_chi}
\end{equation}
with $\epsilon_{esc}$ the fraction of the incoming energy flux that
is taken away from the shock by cosmic rays, and
$G \equiv \frac{3}{2}w + \frac{5}{2}$.\footnote{I assume here that the
non-thermal contributions to the pressure have an adiabatic index of $
\gamma=4/3$.}
One can apply these expressions to the following relation between
plasma temperature and shock temperature:
\begin{equation}
kT_i =  (1 - w)\frac{1}{\chi}\bigl(1 - \frac{1}{\chi} \bigr)V_s^2,
\label{eq_hugoniot3}
\end{equation}
which is a generalized version of Eq.~\ref{eq_hugoniot2},
as can be seen by inserting $w=0$ and $\chi = 4$.
Eq.~\ref{eq_hugoniot3}  and \ref{eq_chi} are graphically depicted as
a function of $w$ and $\epsilon_{esc}$ in Fig.~\ref{fig_hugoniot}.

It is clear from these equations that high resolution spectroscopy
offers the opportunity to learn about the internal
energy budget of the plasma.
Measuring $kT_{\rm p}$ and $V_s$ independently 
gives a handle on cosmic ray acceleration efficiencies. 
This should be combined with a measurement of $kT_{\rm e}$ in order
to test for ion-electron equilibration.
Ideally, one would like to measure 
also the compression factor independently,
in order to alleviate the degeneracy between
$w$ and $\epsilon_{esc}$. On the other hand, it seems unlikely that 
$w < \epsilon_{esc}$, which constrains part of the degeneracy.

So far the interpretation of
most measurements of $kT$ using optical, UV and X-ray spectroscopy
have ignored the role of cosmic ray physics (but see \citep{rakowski08}). 
To some extent this seemed unnecessary,
because spectra were taken from locations from which  
no strong X-ray synchrotron radiation is emitted,
like the northwestern region of SN1006.
The northwestern region of SN1006 has bright \halpha\ emission,
which made it easy to obtain high quality spectra.
The X-ray synchrotron regions, on the other hand, only show weak
\halpha\ emission.
That is unfortunate, since at these synchrotron rims cosmic ray acceleration
is likely to be more efficient. It is not quite clear whether the
lack of \halpha\ emission from the X-ray synchrotron rims is a coincidence.
Three possible reasons for this anti-correlations are 1) the presence
of neutral hydrogen damps plasma wave, decreasing the efficiency
for cosmic ray acceleration \citep{drury96}; 2)
cosmic rays diffusing away from the shock ionize the CSM, supressing
the number of neutral hydrogen atoms entering the shock;
3) bright \halpha\ emission
comes from the densest regions, but as a result of the higher density
the shock velocity has decelerated, thereby diminishing the acceleration
efficiency \citep{vink06d}. 

\subsubsection{Knot g in Tycho's SNR}
A case in point is the \halpha-bright ``Knot g'' in Tycho's SNR. 
The broad \halpha\ line component gives a proton temperature
of $5.9\pm0.9$~keV, indicating a shock velocity of 1700-2200~\kms, depending
on whether ion-electron equilibration is assumed or not.
The average shock velocity of Tycho's SNR is 3100~\kms, based on radio
proper motion studies \citep{reynoso97}, and assuming a distance
of 2.5~kpc. Without cosmic ray acceleration the expected temperature
is $kT_{\rm p}=19$~keV for an unequilibrated plasma, or $kT=11$~keV for an
equilibrated plasma. However, for ``knot g'' the shock velocity
is probably much lower, since the radio proper motion at this
location indicates an expansion parameter $V_s/(R_t/t) \approx 0.25$,
rather than the 0.46 found in the rest of the remnant.
The reason is probably
that the blast wave at ``knot g'' is encountering dense material slowing
down the shock, but also resulting in bright \halpha\ emission.
For the lower expansion velocity of ``knot g'' the measured plasma
temperature is consistent with a shock without cosmic ray modification.
This is at odds with the above mentioned findings that all over this remnant
the contact discontinuity is too close to the forward shock as a result
of cosmic ray acceleration \citep{warren05}: cosmic rays do not seem to
contribute much to the internal energy in ``knot g''.

\subsubsection{The low plasma temperature in LMC SNR 0509-67.5}
One of the fastest expanding young SNRs is 0509-67.5 in the Large Magellanic
Cloud. This remnant has an age of about 400~yr \citep{vink06b,rest05} and
high resolution X-ray spectroscopy with XMM-Newton's RGS experiment
shows line broadenings of $\sigma_v \approx 5000$~\kms,
which is dominated by kinematic Doppler broadening, as the whole
SNR is observed \citep{kosenko08}.
The broadening provides a lower
limit to the shock velocity, since the plasma velocity directly behind
the shock is $3/4V_s$, and most of the line emission comes from
the reverse shock region, which moves probably 30\% slower than
the plasma immediate downstream of the shock velocity. This means
that the shock velocity is at least $V_s \approx 6700$~\kms, but
likely larger than $V_s \approx 8000$~\kms. 
The hydrogen line broadening
has been measured to be $3710\pm 400$~\kms\ (FWHM) \citep{ghavamian07b},
implying a proton temperature of $kT_{\rm p} = 26$~keV.
This translates into a shock velocity 3000-5400~\kms, which is 
much lower than the 
actual shock velocity. The ratio of the measured temperature and the expected
temperature is $\sim 0.2$ (for $V_s \approx 8000$~\kms), 
indicating, according to Fig.~\ref{fig_hugoniot},
either a high partial pressure not coming from the thermal plasma,
$w=0.7$, or a combination of lower $w$ with a relatively
high cosmic ray escape fraction.
Our knowledge of cosmic ray acceleration in 0509-67.5 would be more complete,
if either X-ray synchrotron emission or TeV \gray s would be
detected. Unfortunately, for the moment the remnant is too distant
to identify X-ray synchrotron emission from the shock front.

\subsubsection{The plasma temperature in an 
X-ray synchrotron emitting region in RCW 86}
Ideally one would like to measure post shock temperatures of
region that emit both X-ray synchrotron radiation and/or TeV \gray s.
In the future this can be undertaken with the
next generation of high resolution, imaging X-ray spectrometers,
as are planned for the Japanese NEXT mission or the ESA/NASA/JAXA International
X-ray Observatory (IXO). X-ray and UV spectroscopy have advantages over
\halpha\ spectroscopy, since X-ray line emission does not require a partially
neutral pre-shock gas.

In the mean time one has to concentrate on the faint
\halpha\ line emission coinciding with X-ray synchrotron emission.
In a new study Helder et al. (in preperation) have done that for the
SNR RCW 86, the possible remnant of AD 185 \citep{stephenson02}.
RCW 86 has recently been detected in TeV \gray s by H.E.S.S. (Hoppe et al.
these proceedings). Its X-ray emission is a mix of
X-ray synchrotron , and thermal radiation.
The X-ray emission indicates strong density contrasts, and probably also
strong gradients in shock velocities along the shell
\citep{vink06d}.
\halpha\ emission has been detected all around the shell  \citep{smith97},
but the \halpha\ emission is weak in the northeastern side, where,
the emission is dominated by X-ray synchrotron radiation.
As a consequence ESO's Very Large Telescope is needed to obtain
\halpha\ spectroscopy of this region (Fig.~\ref{fig_rcw86} 
shows the prelimanary
spectrum). Surprisingly, the line width is 1270~\kms (FWHM), corresponding
to a proton temperature of $kT_{\rm p} = 3$~keV, (ignoring
proton-electron equilibrartion).
This is much larger
than in other parts of RCW 86, where typically line width of 500~\kms
are found \citep{ghavamian01}, but it is smaller than expected
for an X-ray synchrotron emitting shock for which
$V_s > 2000$~\kms \citep{aharonian99}.
The plasma temperature appears to be too low,
which can be explained, if cosmic rays have taken up
part of the shock energy. For a reasonable value of 
$V_s \approx 2500$~\kms, Fig.~\ref{fig_hugoniot} suggests $w=0.6$ for
an escape fraction of $\epsilon_{esc}=0$.

However, an alternative scenario to explain the low plasma temperature
in RCW 86 is that the X-ray synchrotron emitting
electrons were accelerated in the past, when the shock velocity was higher.
This requires a long loss time for the electrons, corresponding to
a relatively low magnetic field. Indeed, the value for the magnetic
field derived from the X-ray emitting region is $24\mu$G \citep{vink06d}, 
corresponding to a sufficiently long
loss time of $\sim 500$~yr. 
The new H.E.S.S. results are in agreement with such a low magnetic
field, provided that the TeV emission has a leptonic, rather
than a hadronic origin. 

So, unfortunately, no clear cut answer can be given about the cosmic
ray content of RCW 86, as long as the shock velocity is not
directly measured, or the \gray\ emission process is identified. 
However, RCW 86 illustrates
how seemingly remote areas of astrophysics,
\gray\ astronomy and optical spectroscopy, are both needed to obtain
answers about cosmic ray acceleration efficiencies.

\section{Summary and conclusion}
The topic of these proceedings is \gray\ astrophysics,
a field of astrophysics that is still relatively young, but
has already changed changed our perception of
cosmic ray acceleration in SNRs. 
However, as I have indicated here, more traditional
fields of astronomy valuable clues about cosmic
ray acceleration in young SNRs, as well. 

In reality, not all 
signatures of cosmic ray acceleration are unambiguous. 
For example, the evidence
for high compression ratios is tantalizing \citep{warren05,cassam08}, 
but the high compression ratios all over the remnants,
even those region where cosmic ray acceleration seems not
to be efficient, is confusing, and requires further investigation.
However, it is important that we know now for what kind of multiwavelength 
signatures of cosmic ray acceleration to look, 
and what kind of observations are needed to make further
progress.

Comparing our current knowledge of cosmic ray acceleration
by young SNRs with what was known a decade ago, one sees that
a lot of observational progress has been made.
My guess is that most cosmic ray physicists are a much
more optimistic about the idea that young SNRs are the sources
of cosmic rays up to the ``knee'' than 15 years ago.


\begin{theacknowledgments}
I thank Eveline Helder and Klara Schure
for helpful discussions and providing me with
material for this review. I thank the organizers
for inviting me to this interesting symposium,
and I acknowledge helpful and lively discussions with
Vladimir Zirakashvili, Luke Drury, and Tony Bell.
The author is supported by a Vidi  grant from the 
Netherlands Science Foundation (NWO).
\end{theacknowledgments}



\bibliographystyle{aipproc}   


\def\aj{AJ}
\def\apj{ApJ}
\def\apjs{ApJS}
\def\apjl{ApJ}
\def\pasj{PASJ}
\def\aap{A\&A}
\def\aaps{A\&AS}
\def\nat{Nat}
\def\mnras{MNRAS}
\def\adspr{Adv. Space Research}



\end{document}